\documentclass[aps,prb,showpacs,twocolumn]{revtex4}
\usepackage{epsfig}
\usepackage{amsmath}

\begin{document}

\title{Significant g-factor values of a two-electron ground state in quantum dots 
with spin-orbit coupling}
\author{Yuval Weiss, Moshe Goldstein and Richard Berkovits}
\affiliation{The Minerva Center, Department of Physics, Bar-Ilan University,
  Ramat-Gan 52900, Israel}

\begin{abstract}

The magnetization of semiconductor quantum dots in the presence of spin-orbit 
coupling and interactions is investigated numerically. When the dot is occupied 
by two electrons we find that a level crossing between the two lowest many-body 
eigenstates may occur as a function of the spin-orbit coupling strength. This 
level crossing is accompanied by a non-vanishing magnetization of the ground-state.
Using first order perturbation theory as well as exact numerical diagonalization of
small clusters we show that the tendency of interactions to
cause Stoner-like instability is enhanced by the SO coupling. The resulting g-factor 
can have a significant value, and thus may influence g-factor measurements. 
Finally we propose an experimental method by which the predicted phenomenon can be observed.

\end{abstract}

\pacs{73.21.La,71.10.Pm,75.75.+a}

\maketitle

\section{Introduction}
The effect of spin-orbit (SO) coupling on the energy spectrum of quantum dots (QDs) 
and metallic grains
has attracted notable attention in the recent years \cite{halperin86,beenakker97,alhassid00}.
Much experimental and theoretical effort has concentrated
on the magnetization of mesoscopic samples. For example, measurements of the g-factors of 
nano-particles using tunneling spectroscopy
\cite{ralph95,davidovic99,petta01} have led to several theoretical studies\cite{brouwer00,matveev00} 
which treated the electrons in the quantum dots as non-interacting particles.
Other theoretical studies have considered interactions as well, while investigating 
the interplay between interactions and disorder in quantum dots without SO coupling. 
It was shown that the combination
of these effects can lead to non-trivial spin polarization 
\cite{berkovits98,brouwer99,berkovits99,kurland01,benenti01,hirose02,usaj02}.
In this paper we present an additional mechanism which can lead to spin polarization
and non-trivial g-factor, where the role typically played by disorder is taken
by the SO coupling.

Usually the g-factor is defined through the splitting of the Kramers doublets \cite{kramers30,Merzbacher} 
in the presence of a weak magnetic field. Namely, the g-factor of the $i$-th single-particle level
with spin $\sigma$ is given by
\begin{eqnarray} \label{eqn:g_gdef}
g_{i,\sigma} = \frac{2 \left[ \epsilon_{i\sigma}^{(0)}-\epsilon_{i\sigma}^{(H)} \right]}{\mu_B H},
\end{eqnarray}
where $\epsilon_{i\sigma}^{(H)}$ ($\epsilon_{i\sigma}^{(0)}$) is the corresponding energy level 
in the presence (absence) of a weak magnetic field $H$
and $\mu_B$ is the Bohr magneton. The spin index $\sigma \in \{+,-\}$ is used to 
denote the two time-reversed states according to the sign of the $z$ component of their average magnetic moments.
In the absence of a magnetic field each level is two-fold degenerate,
and this degeneracy is lifted by the magnetic field, which increases the energy of one of 
the levels and decreases the energy of the other. Therefore, $g_{i,\sigma}$ as defined by this formula can have either
sign, depending on the direction of the energy change. The ground-state energy always decreases
when a magnetic field is applied, thus the g-factor of the ground-state obtained by Eq.~(\ref{eqn:g_gdef}) 
is positive. Usually, the value of $g$ does not depend on the spin index, at least
to zeroth order in $H$, so that one can denote the g-factor of the $i$-th level as $\pm g_i$,
with the convention that $g_i \ge 0$.

For free electrons the g-factor is constant, $g_{i}=2$ for each level $i$, 
and this value is more or less correct also for bulk measurements in various 
metals \cite{halperin86}. However, in experiments performed on metallic nano-particles, 
values which are significantly less than the free value of the g-factor were 
obtained\cite{davidovic99,petta01}. Moreover, large fluctuations in the measured 
values were seen. These findings attracted much theoretical 
attention, and resulted in studies which have obtained, within the framework of the 
random matrix theory (RMT), a description of the g-factor probability distribution in 
the presence of SO coupling and disorder but in the absence of interactions \cite{brouwer00,matveev00}. 
In a recent work, the statistical properties of these distribution functions 
were related to several physical observables \cite{mucciolo06}. 
According to these results, the SO coupling influences the probability distribution 
of the g-factors of the discrete energy levels. The distribution function was 
shown to be universal, where the width is expressed in terms of various physical 
parameters. The presence of strong SO coupling and disorder results in 
sample to sample fluctuations of the g-factor. 
Moreover, the g-factor is expected to fluctuate also between 
different levels of a specific sample, with a distribution 
function determined by RMT. 

Indeed, recent measurements of nano-particles have obtained g-factors  
which seem to be consistent with RMT predictions.
For example, several experimental studies of metallic three dimensional 
nano-particles have shown the reduction of the measured g-factor as a function of 
the spin-orbit coupling strength.
For aluminum nano-particles, in which the SO coupling is negligible, the measured 
g-factor values are approximately those of free electrons \cite{ralph95} (i.e., $g \approx 2$), 
while for gold nano-particles, in which the SO coupling is strong, the 
measured g-factors were in the range of $0.28-0.45$.\cite{davidovic99}
Furthermore, by extracting several g-factors from each sample, Petta and Ralph 
have succeeded to present an impressive confirmation of the theoretical 
RMT distribution function \cite{petta01}. 

Nevertheless, according to Eq.~(\ref{eqn:g_gdef}) the g-factor measurement should compare 
the specific single-particle energy level before and after the magnetic field is applied. 
However, practical experiments usually differ from that approach in two points. 
First, measurements are usually related to the total energy of the system, 
and not to that of a specific level. Second, the measurement of the energy is 
sometimes indirect, as is the case in tunneling spectroscopy.

These two points can be ignored 
if one neglects the interaction between particles. For a non-interacting system with 
an odd number of electrons, $n_e=2p+1$, the change of the total ground-state energy due to 
magnetic field is equivalent to that of the highest filled level. 
$n_e-1$ electrons populate $p$ Kramers pairs, where in each pair one level 
increases and the other decreases in the presence of a magnetic field, so that their
total contribution vanishes. 
The only contribution to the g-factor comes from the single electron occupying 
one level of pair $p+1$,
so that if we define the g-factor of the ground-state with $n_e$ electrons by
\begin{eqnarray} \label{eqn:g_gdef2}
g(n_e) = \frac{2 \left[ E_{gs}^{(0)}(n_e)-E_{gs}^{(H)}(n_e) \right]}{\mu_B H},
\end{eqnarray}
where $E_{gs}^{(H)}(n_e)$ represents the total ground-state energy in the presence 
of a magnetic field $H$, then $g(2p+1)=g_{p+1}$.
In addition, when the number of electrons $n_e$ is even, 
the total ground-state energy is not expected to
change when a magnetic field is applied, since all the filled levels divide into pairs, in which 
the movement of one level is compensated by the other (to first order in $H$).
Therefore, for an even number of electrons, 
a calculation of the g-factor using Eq.~(\ref{eqn:g_gdef2}) gives $g(2p)=0$. 

The second point, regarding the indirect energy measurement, requires an interpretation 
of the experimental results.
For example, using tunneling spectroscopy one measures the 
gate voltage value for which a conductance peak of a QD occurs. At such an event 
the energies of the QD with $n_e-1$ and $n_e$ electrons and the gate voltage 
$V_g$ are related by the equation $eV_g = E_{gs}(n_e)-E_{gs}(n_e-1)$.
When a magnetic field is applied, the position of the peak will change as a function of $H$. Therefore, 
by denoting the measured g-factor by $\tilde g$, one can analyze the
peak motion in order to determine the g-factor, by calculating
\begin{eqnarray} \label{eqn:g_gdef_tot}
\tilde g &=& \frac{2 \left[ eV_g(0)-eV_g(H) \right]}{\mu_B H} 
= g(n_e) - g(n_e-1).
\end{eqnarray}
Since either $n_e$ or $n_e-1$ is even, its corresponding g-factor vanishes, and 
thus $\tilde g$ is equivalent to the other g-factor. Namely, $\tilde g = g(n_e)$ 
or $\tilde g = -g(n_e-1)$. Actually, since each peak is split in the presence 
of a magnetic field into two peaks having an opposite magnetic field dependence, 
extracting $\tilde g$ from successive peaks results in the set of the 
single-particle g-factors, i.e. $g_1, -g_1, g_2, -g_2, \dots$. 

As mentioned above, measurements done using tunneling spectroscopy have indeed
obtained g-factors which can be interpreted using RMT predictions.
Nevertheless, as we have discussed, Eq.~(\ref{eqn:g_gdef2}) is equivalent 
to Eq.~(\ref{eqn:g_gdef}) only for systems of non-interacting particles.
Once interactions between electrons are important, it should be emphasized 
that Eq.~(\ref{eqn:g_gdef2}) is a definition of a {\bf many-particle g-factor}, 
which depends on the total magnetization of the ground-state wave-function. For 
example, one can obtain spin contribution to the g-factor which is larger than $2$, a 
phenomenon that cannot happen for a single-particle 
g-factor.\cite{berkovits98,brouwer99,berkovits99,kurland01,benenti01,hirose02,usaj02}
Indeed, by adding an interaction term to the RMT Hamiltonian, an increase of the 
g-factor fluctuations was reported.\cite{gorokhov03,gorokhov04} It was shown that 
the interactions result in a possibility of getting non-trivial spin values in the 
ground-state, and accordingly in an optional enhancement of the g-factor 
to values greater than $2$.

Although the theoretical studies of Refs.~\onlinecite{gorokhov03} and \onlinecite{gorokhov04} 
were performed for an odd-electron occupation, their results suggest the possibility 
of a non-trivial spin polarization for the even-electron case as well. 
If, for any reason the g-factor of an even-electron ground-state 
indeed differs from zero, then \emph {the quantity measured in tunneling spectroscopy may 
not equal the single-level g-factor nor the many-particle g-factor}. In such a 
case it should be related to the difference between two 
many-particle g-factors, as shown in Eq.~(\ref{eqn:g_gdef_tot})

In principle, the above description of the g-factor holds for both metallic and 
semiconducting dots. However, in semiconducting dots the strength of the SO coupling 
can be tuned by use of a gate voltage\cite{nitta97,engels97}.
Furthermore, several significant implications of SO effects in semiconductors, such as
spin-polarized field effect transistor\cite{datta90} and spin Hall 
effect\cite{kato04,wunderlich05}, have recently attracted notable attention. 

With this in mind, we investigate in the current paper the ground-state magnetization 
properties of semiconducting QDs with SO coupling where interactions between 
the electrons are considered. We show that the interplay between the SO coupling and the 
electron-electron interactions may result in a level crossing (LC) between the two lowest
many-body levels. When these states are close in energy, the magnetic field splits them into 
two polarized states with a finite magnetization. As a result, there is a possibility to have 
a significant g-factor in the two-particle ground-state.
Finally, we propose an experimental method which can be used in order to
observe the predicted phenomenon.

We note that we have neglected so far in the introduction
the orbital effect and its influence
on the magnetization of nano-particles. For three-dimensional (3D) nano-particles
this is reasonable\cite{matveev00}.
On the other hand, for two-dimensional (2D) systems the orbital effect is expected
to play an important role. 
For example, due to the orbital effect, the single-particle g-factor
can exceed the value of $2$. In addition, the g-factors of two 
levels belonging to the same Kramers pair, $g_{i,+}$ and $g_{i,-}$, might be different.
As a result, the g-factor of the doubly occupied ground state, which can be simply 
written, when the electrons are non-interacting, as
$g(2) = g_{1,+} + g_{1,-}$, may not vanish. However, this contribution to $g(2)$
which is the quantity of interest in this work,
is linear in the magnetic field $H$, and thus can be neglected for the weak fields
used in such measurements.

The rest of the paper is organized as follows. In the next section we describe
the model Hamiltonian we use in order to incorporate, beside the magnetic field, both 
SO coupling and interactions between electrons. In section \ref{sec:g_I0} we present 
results for a system with non-interacting particles, 
which are shown to reproduce some known ground-state properties. 
In addition, we find that there are specific values of the SO coupling strength, in which the Kramers 
doublet remains degenerate even when a magnetic field is applied. 
The effects of interplay between SO and electron-electron interactions are considered
in section \ref{sec:g_I}. Our results point out that a finite
magnetization can be obtained for systems with an even-particle occupancy. In section
\ref{sec:g_phys} we discuss the experimental relevance of this finding, i.e., the
possibility that it might affect practical g-factor measurements. 

\section{Model}
In order to model the semiconducting QD we use a tight-binding description of a finite 2D
lattice with $A$ columns and $B$ rows (the number of sites is denoted by $N=AB$), with open
boundary conditions, which is occupied by $n_e$ electrons. 
As a result of a coupling between the spin degree of freedom
and the orbital motion a finite probability for spin-flips during hopping processes exists. 
Separating the interactions from the free part, one can write the Hamiltonian as 
$\hat H_{\rm QD} = \hat H_0 + \hat H_{\rm int}$, 
where the free part in the absence of disorder
can be divided to a hopping term and a Zeeman term, i.e.
$\hat H_0 = \hat H_{\rm hop}+ \hat H_B$.
The hopping part of the Hamiltonian is
\begin{eqnarray} \label{eqn:g_Hdot_0}
\hat H_{\rm hop} =
- \sum_{m,n,\sigma,\sigma^\prime} ( &V_{x}& \hat a^\dagger_{m,n,\sigma} \hat a_{m,n+1,\sigma^\prime} \\ \nonumber
+ &V_{y}& \hat a^\dagger_{m,n,\sigma} \hat a_{m+1,n,\sigma^\prime} + H.c.),
\end{eqnarray}
where $\hat a^\dagger_{m,n,\sigma}$ ($\hat a_{m,n,\sigma}$)
is a creation (annihilation) operator of an electron with spin $\sigma$ in the
lattice site placed in row $m$ and column $n$. The matrices $V_x$ and $V_y$ are 
given by the Ando model\cite{ando89}, which is the discrete version for
the Rashba spin-orbit coupling\cite{rashba}, as
\begin{equation} \label{eqn:Intro_Hso_V}
V_x = \left( \begin{array}{cc}
	V_1 & V_2 \\
	-V_2 & V_1 \\
	\end{array} \right)~~;~~~
	V_y = \left( \begin{array}{cc}
	V_1 & -iV_2 \\
	-iV_2 & V_1 \\
	\end{array} \right)~,
\end{equation}
where $V_1$ ($V_2$) is the hopping matrix element, for events which conserve (flip) the spin.
The overall hopping amplitude, 
$t = \sqrt{V_1^2 + V_2^2}$, is taken as the energy unit of the problem. 
In other words, all energies are expressed in terms of $t$.

The strength of the SO coupling can be expressed by the ratio between 
the absolute value of the spin-flip amplitude and that of the total hopping element.
Using a dimensionless parameter 
$\lambda = \frac {V_2}{\sqrt{V_1^2+V_2^2}}= V_2/t$, we examine the entire range of $\lambda$, between 
very weak ($\lambda \rightarrow 0$) and very strong ($\lambda \lesssim 1$) spin-orbit 
coupling. Realistic values for $\lambda$ are between $0$ and $0.5$.\cite{ando89,ando82}
As mentioned above, these values can be controlled by tuning the 
gate voltage\cite{nitta97,engels97}.

We now add a perpendicular magnetic field to our 2D sample,
and we choose a gauge in which the vector potential is $A=-Hy \hat x$. 
The Zeeman term in the Hamiltonian is thus diagonal
in spin space, and can be written as
\begin{eqnarray} \label{eqn:g_Hdot_B}
\hat H_B = \displaystyle \mu_B H \sum_{m,n,\sigma} \sigma \hat a^\dagger_{m,n,\sigma} \hat a_{m,n,\sigma},
\end{eqnarray}
where $\sigma = \pm 1$. 

With the gauge chosen, one has to modify the hopping elements in the $\hat x$ direction, 
according to the Peierls substitution\cite{peierls}, and write 
$V_x \rightarrow V_x e^{-i \theta m}$. In this expression $m$ is the row number
and $\theta$ is a phase, that can be written as 
$\theta = \frac {2 \pi H s^2}{\phi_0}$, where 
$s$ is the lattice constant and $\phi_0 = hc/e$ is the magnetic flux quantum.
Thus, $\theta$ is a dimensionless parameter, that measures the magnetic flux
throughout a lattice unit cell, in units of the quantum flux $\phi_0$.

The Zeeman energy can be related to the hopping phase $\theta$ and to the hopping amplitude $t$
by the following considerations. 
One can express the absolute value of the Zeeman energy as
$\mu_B H = \mu_B \phi_0 \frac {\theta} {2 \pi s^2}$. Substituting the physical constants
$\mu_B \phi_0 = \frac {\pi \hbar^2}{m_0}$, where $m_0$ is the electron mass,
and using the relation $t=\frac{\hbar^2}{2m_{\rm eff}s^2}$,
where $m_{\rm eff}$ is the effective mass at the bottom of the band,
one gets $\mu_B H = \frac {\theta \hbar^2} {2 m_0 s^2} = \theta t \frac{m_{\rm eff}}{m_0}$.
The factor $\frac{m_{\rm eff}}{m_0}$ depends on the specific type of the QD,
and in general, $m_{\rm eff} \approx m_0$ for metallic grains while $m_{\rm eff} < m_0$
for semiconducting ones. In the current study we use 2D geometry which is suitable for modeling typical
semiconducting QDs. Moreover, the Ando model which incorporates the spin-orbit coupling was
originally proposed
for surfaces of III-V compound semiconductors \cite{ando89}. Thus we set for the rest of
this paper $\frac{m_{\rm eff}}{m_0} \approx \frac{1}{15}$, as in the case of GaAs.
However, we have checked that tuning this value does not lead to 
a qualitative change of the main results.
Finally, since all energies are measured in units of $t$, the strength of the Zeeman term
$\mu_B H / t$ determines exactly the hopping phase.

At last, the interactions term in the Hamiltonian is
\begin{eqnarray} \label{eqn:g_Hdot_int}
\hat H_{\rm int} = U \displaystyle \sum_{m,n}
{\hat a}^{\dagger}_{m,n,\uparrow}{\hat a}_{m,n,\uparrow}
{\hat a}^\dagger_{m,n,\downarrow}{\hat a}_{m,n,\downarrow},
\end{eqnarray}
which represents a Hubbard interaction with strength $U$. 

The Hamiltonian $\hat H_{\rm QD}$ is exactly diagonalized using the Lanczos
procedure, for lattices of up to $15 \times 14$ sites, occupied by $1$ or $2$ electrons,
and its lowest eigenstates are numerically found.
In order to calculate the spin polarization of the QD we apply a weak magnetic field 
along the $\hat z$ axis and calculate the expectation value of $\hat S_z$ for the lowest levels. 
For the g-factor calculations, we compare the ground-state energies with and without the magnetic 
field for each sample, and use Eq.~(\ref{eqn:g_gdef2}).
The strength of the magnetic field we apply is $\mu_B H/t \sim 10^{-4} - 10^{-3}$, 
and for an experimental system in which the mean level spacing is $0.1-1 ~ meV$, it
is equivalent to a magnetic field of $10-1000 ~ G$, in correspondence with realistic measurements.

\section{Non-interacting electrons}
\label{sec:g_I0}
\begin{figure}[htbp]
\vskip 1truecm
\centering
\includegraphics[width=3in,height=!]{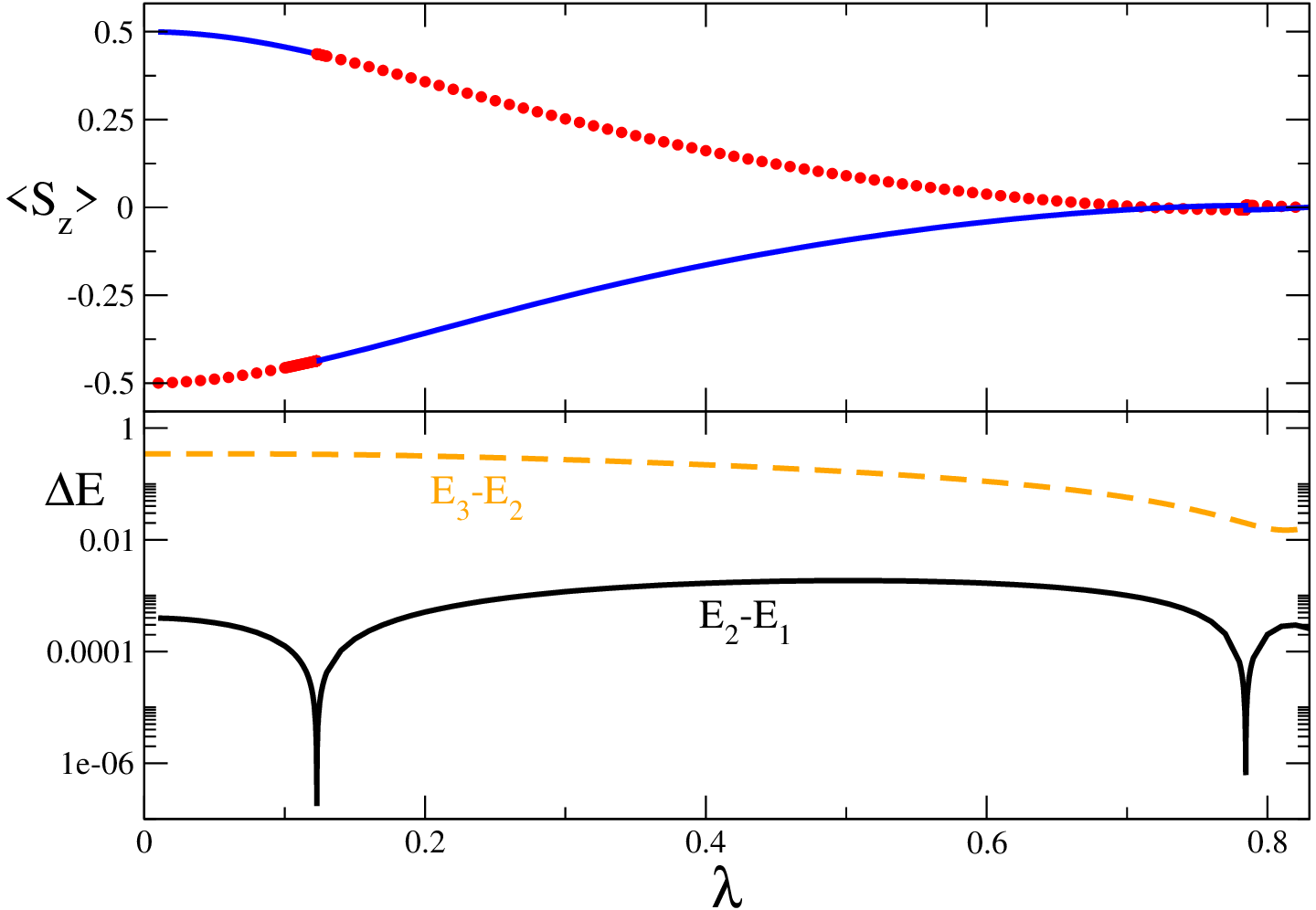}
\caption[Single-level Sz of non-interacting electrons]
{\label{fig:sz_n1} 
(Color online) 
$\langle S_z\rangle$ of the lowest two single-particle levels (upper panel) and $\Delta E$, the 
energy difference between them (lower panel, notice the semi-logarithmic scale), 
calculated for a system of $8 \times 7$ sites in the presence of a magnetic field, 
as a function of the SO coupling strength.
The value of $\langle S_z\rangle$ switches abruptly between the two levels 
(one level is shown by symbols and the other by a line)
near $\lambda=0.12$, where the energy difference vanishes, implying
a level crossing between the lowest two levels. The energy of the third level, however, 
remains much higher (lower panel, dashed line).
}
\end{figure}

We start with non-interacting particles, by taking $U=0$. 
Without the magnetic field, all single-particle states (and in 
particular the ground-state) are doubly-degenerate (the Kramers degeneracy) \cite{kramers30,Merzbacher}. 
When a magnetic field is applied, it splits this degeneracy, and one gets to zeroth order
in the magnetic field, $\langle S_z^{(1)}\rangle = - \langle S_z^{(2)}\rangle$,
where $\langle S_z^{(m)} \rangle$ denotes the expectation value of the
operator $\hat S_z$ in the $m$-th eigenfunction ($m=1$ being the ground-state).
For $\lambda \rightarrow 0$, $\left| \langle S_z\rangle \right| \rightarrow \frac{1}{2}$.
When the SO coupling increases, a general decrease of 
$\left| \langle S_z\rangle \right|$ 
can be expected, and this trend can be seen
in the upper panel of Fig.~\ref{fig:sz_n1}.

However, one can see that the value of $\langle S_z\rangle$ switches abruptly
between these two levels near $\lambda=0.12$. This is a sign of a level crossing (LC),
which can be seen by looking at the energy difference between these levels
(lower panel of Fig.~\ref{fig:sz_n1}). The switching of $\langle S_z\rangle$
occurs exactly when the energy difference vanishes. 
We note that such crossings occur also for large values of $\lambda$.



It is important to notice that the LC presented here occurs
between states which belong to the same Kramers pair (i.e., 
states which are the time reversal of each other).
The energy difference between states of different pairs is much larger,
and although it is reduced by the SO coupling, yet it 
is usually much larger than the contribution of the weak magnetic field we apply to the energy 
(see the dashed line in the lower panel of Fig.~\ref{fig:sz_n1}).
As a result, crossings between states which originate from different pairs
are much less probable.

In the next section we study the case of doubly-occupied systems. For non-interacting 
electrons, based on the results of the current section, it is clear that a LC between
the lowest two doubly-occupied states is improbable, since their Slater determinants 
contain single-particle states from different Kramers pairs.
Nevertheless, as will be shown in the next section, the electron-electron interaction
can change this picture qualitatively. 

\section{Interplay between Interactions and Spin-Orbit Coupling}
\label{sec:g_I}

\begin{figure}[htbp]
\vskip 1truecm
\centering
\includegraphics[width=3in,height=!]{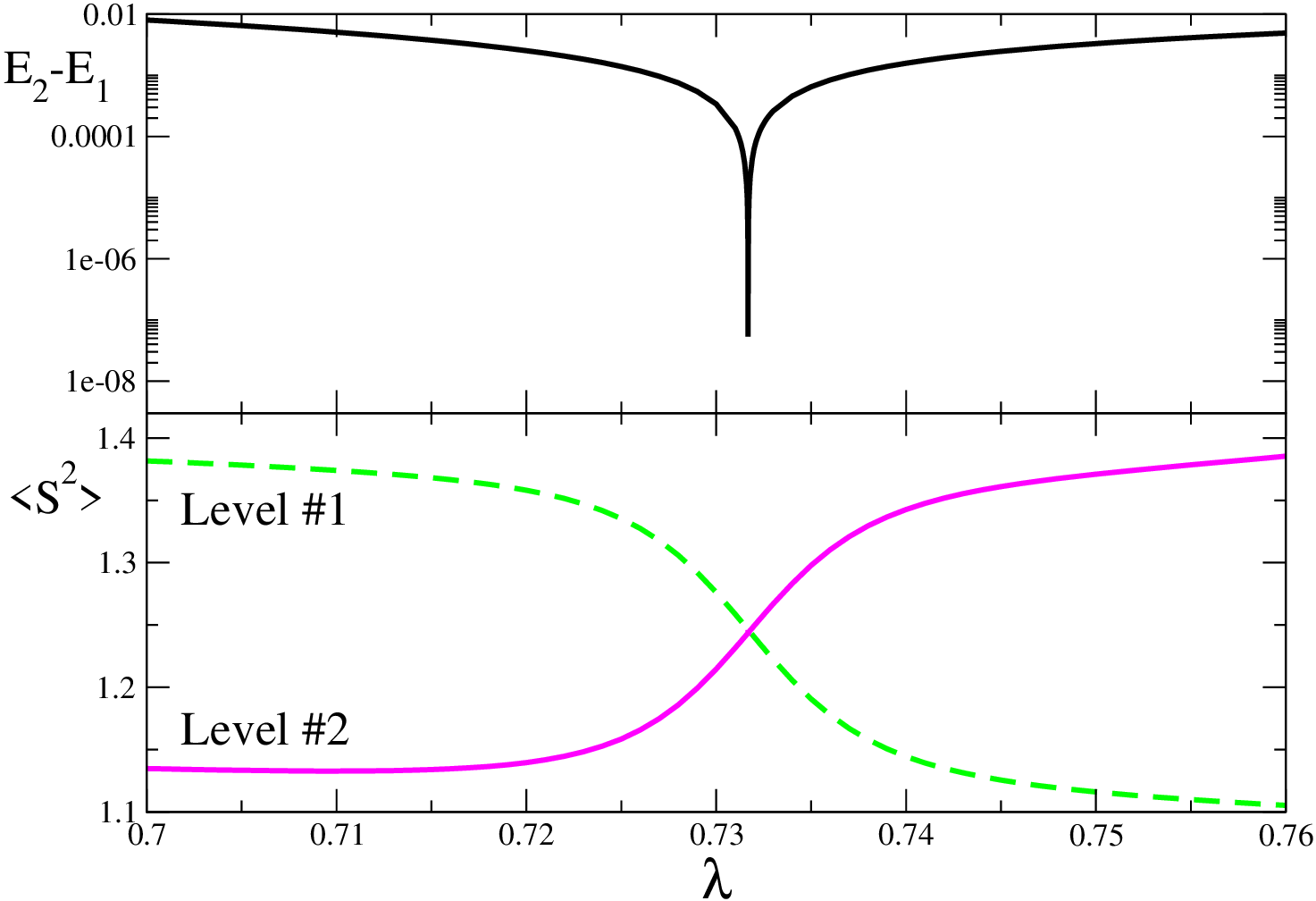}
\caption[Level crossing of the doubly-occupied states with interaction]
{\label{fig:dE_S2_n2} 
(Color online) Typical results of the level crossing of the two lowest doubly-occupied states. The results 
shown were obtained for a system of $8 \times 7$ sites, with $U=3t$. 
Upper panel: the energy difference $E_2 - E_1$ 
is shown as a function of the spin-orbit coupling strength $\lambda$ 
(notice the semi-logarithmic scale). The dip shows the crossing point.
Lower panel: the switch of $\langle \hat S^2 \rangle$ 
between these two states, which is centered at the same place.
}
\end{figure}

We now turn to study the effect of interactions
on the behavior of the g-factor in the presence of SO. 
Calculating the ground-state energies of the two lowest doubly-occupied many-body states, one finds that
there is a LC between these states, at a certain value of the SO coupling (denoted in the following by $\lambda_c$),
as can be seen 
in the upper panel of Fig.~\ref{fig:dE_S2_n2}. 
In the vicinity of $\lambda_c$, the expectation value of $\hat S^2$ switches smoothly
between these states, as is shown in the lower panel of Fig.~\ref{fig:dE_S2_n2}.

\begin{figure}[htbp]
\vskip 1truecm
\centering
\includegraphics[width=3in,height=!]{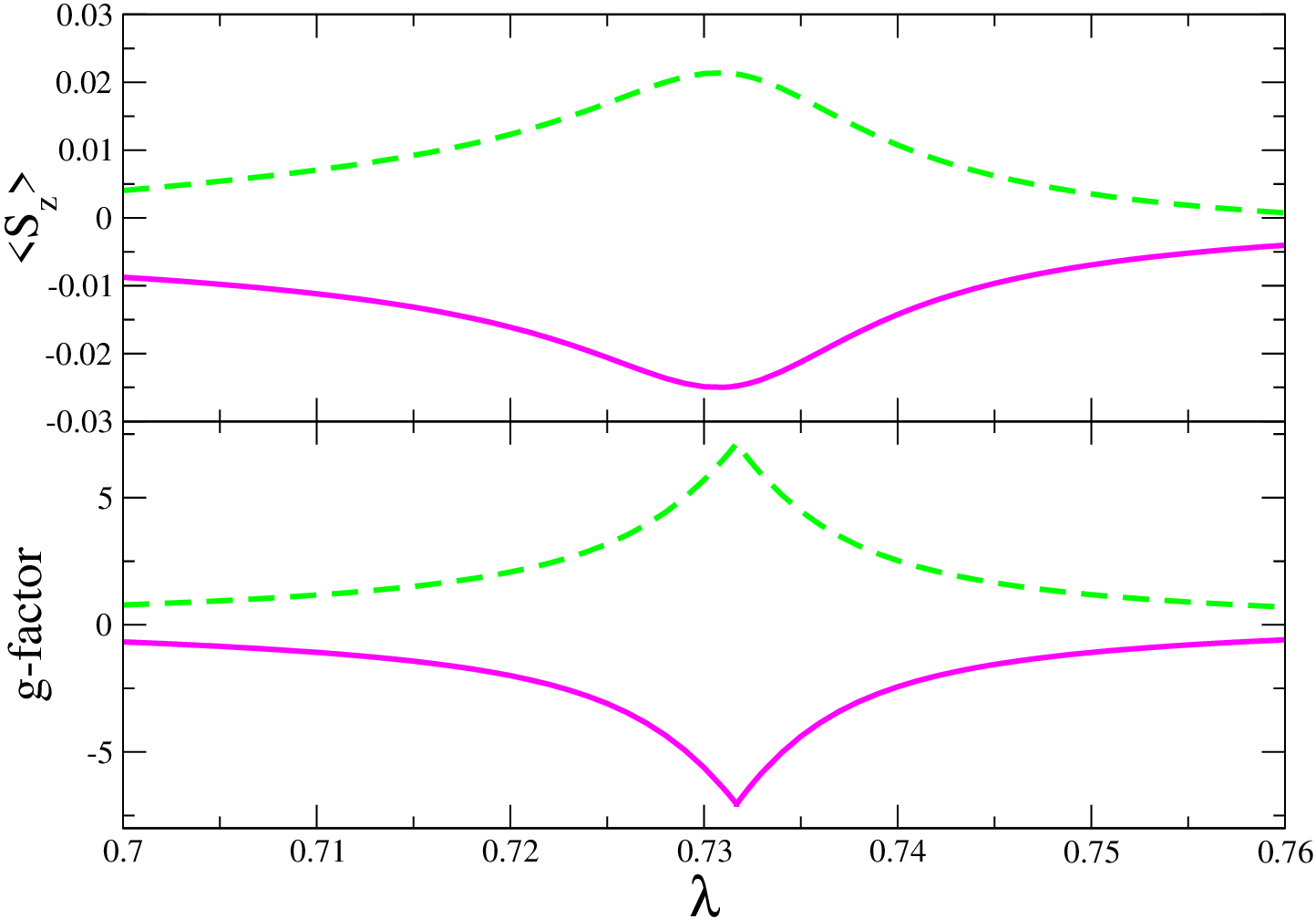}
\caption[Spin polarization and g-factors in the level crossing regime]
{\label{fig:Sz_n2_I} 
(Color online) Typical results of the spin polarization $\langle \hat S_z \rangle$ (upper panel) 
and the g-factor, calculated using Eq.~(\ref{eqn:g_gdef2}) (lower panel),
of the two lowest doubly-occupied states, in the regime of the level crossing between them. 
The results shown were obtained for a system of $8 \times 7$ sites, with $U=3t$
and $\mu_B H=10^{-4} t$. 
}
\end{figure}

As noted in the previous section, such a LC does not exist for non-interacting two-electron
states since it involves levels belonging to different single-particle Kramers pairs. 
Moreover, in the cases when there is a LC in the non-interacting system,
i.e., between single-particle levels belonging to the same Kramers pair, the g-factor 
vanishes at the LC point (to zeroth order of $H$; it has however a linear 
magnetic field dependence from orbital effects).
On the other hand, in the case of interacting electrons we find that both states 
involved in the LC have a significant magnetization of zeroth order in $H$.
The magnetization properties, i.e., $\langle \hat S_z \rangle$ and the g-factor,
do not present a smooth switching as for $\langle \hat S^2 \rangle$ 
in the vicinity of the LC. Instead, 
when the energies of the two states become close enough to each other so that 
the energy associated with the magnetic field becomes important, 
both states develop a spin polarization as can be seen in Fig.~\ref{fig:Sz_n2_I}.
This leads to an enhancement of
$\langle \hat S_z \rangle$ in the crossing region, and to significant values
of the g-factor.

The significant values obtained for the g-factor are crucially related to the
degeneracy point (the LC). 
Far from this point, when the two lowest many-particle states are not degenerate, 
each of these states $\psi_1$ and $\psi_2$ is the time reversal of itself, 
i.e., $\mathcal{T}(\psi_1)=\psi_1$ and $\mathcal{T}(\psi_2)=\psi_2$,
where $\mathcal{T}$ is the time reversal operator.
This immediately implies $\langle \hat S_z \rangle = 0$ for both states,
and the corresponding g-factors vanish as well. This picture changes in the
vicinity of the degeneracy point, where the magnetic field breaks the degeneracy by 
polarizing both states. This of course results in a finite value of the g-factor. 
Such non-vanishing g-factor values can thus be seen as long as the energy 
associated with the magnetic field is larger than the level separation.
Accordingly, as the magnetic field is enhanced, the peaks in 
$\langle \hat S_z \rangle$ and $g$ get wider.

\begin{figure}[htbp]
\vskip 1truecm
\centering
\includegraphics[width=3in,height=!]{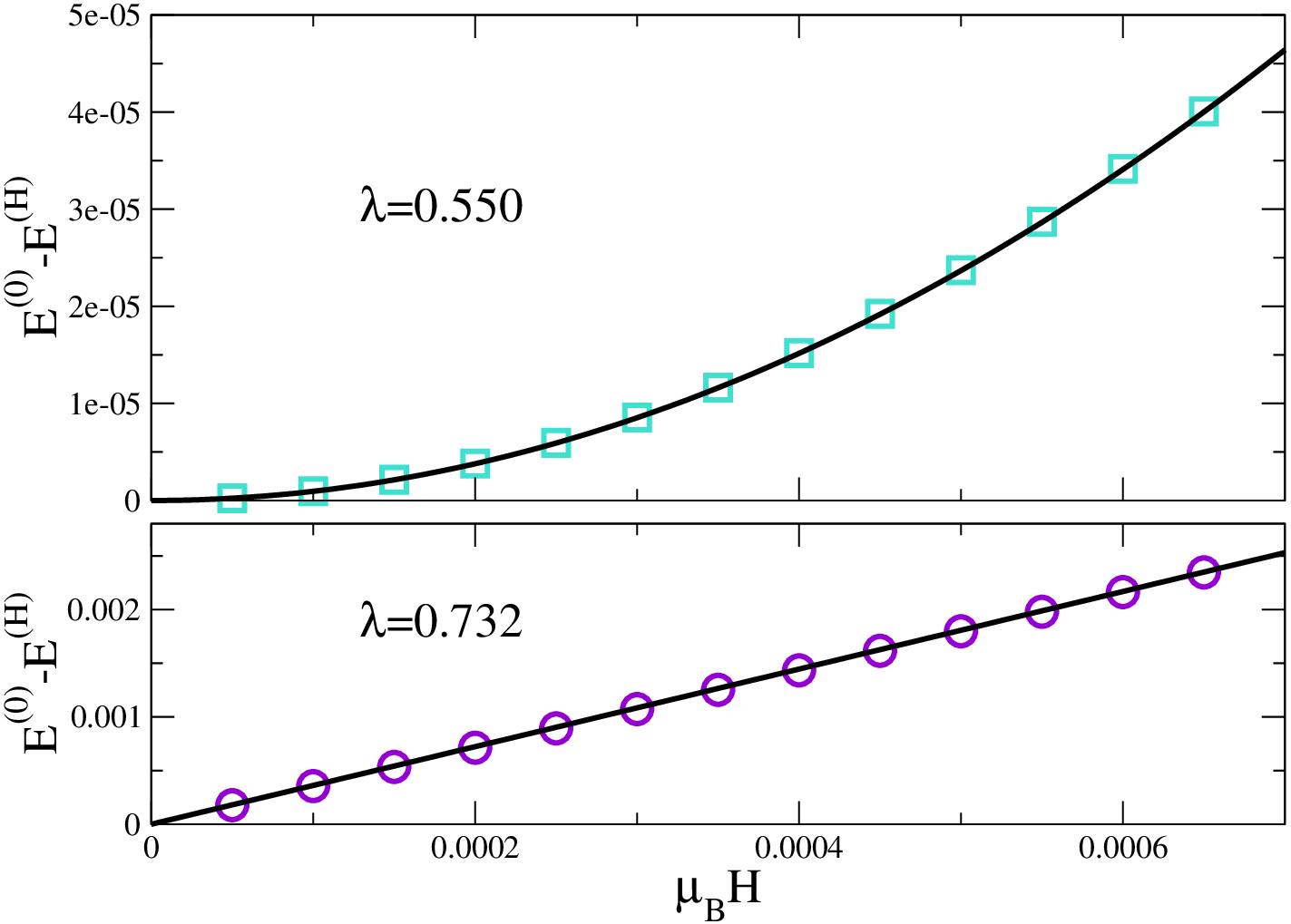}
\caption[Dependence of the energy on the magnetic field in the level crossing regime]
{\label{fig:E_H_n2_I} 
(Color online) The dependence of the energy on the magnetic field is compared between the regime of a
level crossing (lower panel), to another arbitrary point (upper panel). The results shown by symbols were obtained 
for a system of $8 \times 7$ sites, with $U=3t$, and the solid lines represent quadratic
(upper panel) and linear (lower panel) fits. 
}
\end{figure}

The dependence of the energy on the magnetic field is shown in 
Fig.~\ref{fig:E_H_n2_I}, with a comparison between the LC regime to
an arbitrary point. In the latter, a quadratic dependence of the 
ground-state energy on the magnetic field is clearly seen.
On the other hand, near the LC point each of these states has a significant 
magnetization, and the dependence of the energy on the magnetic field is linear, 
with a finite value of the g-factor.

\begin{figure}[htbp]
\vskip 1truecm
\centering
\includegraphics[width=3in,height=!]{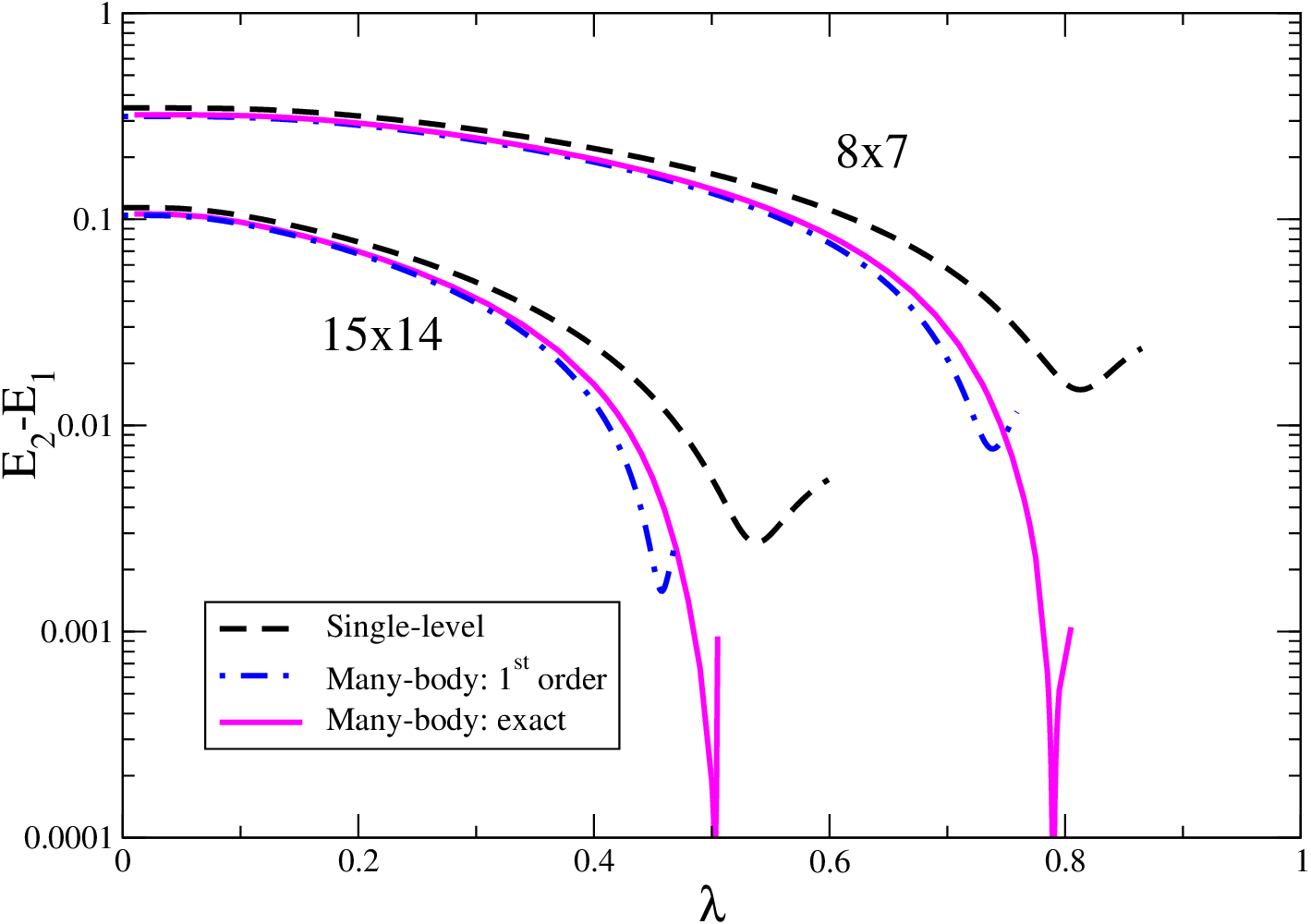}
\caption[Dependence of the energy on the magnetic field in the level crossing regime]
{\label{fig:dE_dpt} 
(Color online) The dependence of the energy difference between the two lowest many-body levels on 
the SO strength for systems of $8 \times 7$ and $15 \times 14$ sites. The results shown are of the
non-interacting case (dashed curve), and of interacting electrons with $U=t$
calculated exactly (solid line) or by using first order perturbation theory (dashed-dotted line).
notice the semi-logarithmic scale.
}
\end{figure}

These phenomena can be given a simple interpretation.
Kinetic energy considerations make it advantageous to put the two electrons in the
same orbital level, and create an unpolarized ground state. Repulsive interactions, however,
cause a polarized ground state to be preferred, since the Pauli principle then tends to separate
the electrons. Usually, the kinetic energy wins. However, SO coupling tends to reduce
the single-particle level spacing 
so at some point the
interactions win, and a Stoner-like instability emerges\cite{stoner47,kurland00}.
In order to support this intuitive picture,
we have calculated the energy difference between the two lowest many-body levels using
first order perturbation theory in the interaction strength. As can be seen in Fig.~\ref{fig:dE_dpt},
the interaction tends to decrease the energy difference between the lowest 
two many-body levels. When the electrons are non-interacting, the levels 
approach each other with increasing SO coupling, yet the minimal distance 
between them is much larger than the magnetic energy. The presence of
interactions enhances this tendency, towards the situation in which a LC is
possible.

From these results one can conclude that whereas the g-factor of a doubly-occupied
system can be neglected for most values of $\lambda$, it nevertheless has a significant 
value near $\lambda_c$. As is shown in Fig.~\ref{fig:dE_dpt},
when the system size increases the instability and the g-factor peak occurs for smaller
values of the SO coupling. In the next section we argue that 
such g-factor values might be significant even for realistic sample sizes
and physical parameters, and thus they should not be neglected 
when analyzing experimental data.

\section{Experimental Relevance and Discussion}
\label{sec:g_phys}
In order to check whether the g-factor peak presented in the previous section occurs 
for realistic systems, one must study how the system size modifies this behavior. 
When the system is enlarged, one must be careful to leave the other physical 
parameters unchanged. The strength of the interactions is usually described by
the parameter $r_s$, which is defined through the ratio between the potential
and the kinetic energies. The kinetic energy per electron in 2D samples goes like
the electron density $n=n_e/N$. For Coulomb interactions one has
$E_p \sim n \iint U_C/r \,dx \,dy$ per electron, where $U_C$ is the Coulomb interaction strength 
between sites separated by one lattice constant. Since $r \sim n^{-1/2}$
one gets $E_p \sim U_C \sqrt{n}$ and thus $r_s \sim U_C/\sqrt{n}$.
However, for Hubbard interactions $E_p \sim n \iint U \delta(x-x_0) \delta(y-y_0) \,dx \,dy = nU$,
so that $r_s \sim U$.
Therefore, in order to keep $r_s$ constant, the value of $U$ should stay unchanged
when the system size increases.

\begin{figure}[htbp]
\vskip 1truecm
\centering
\includegraphics[width=3in,height=!]{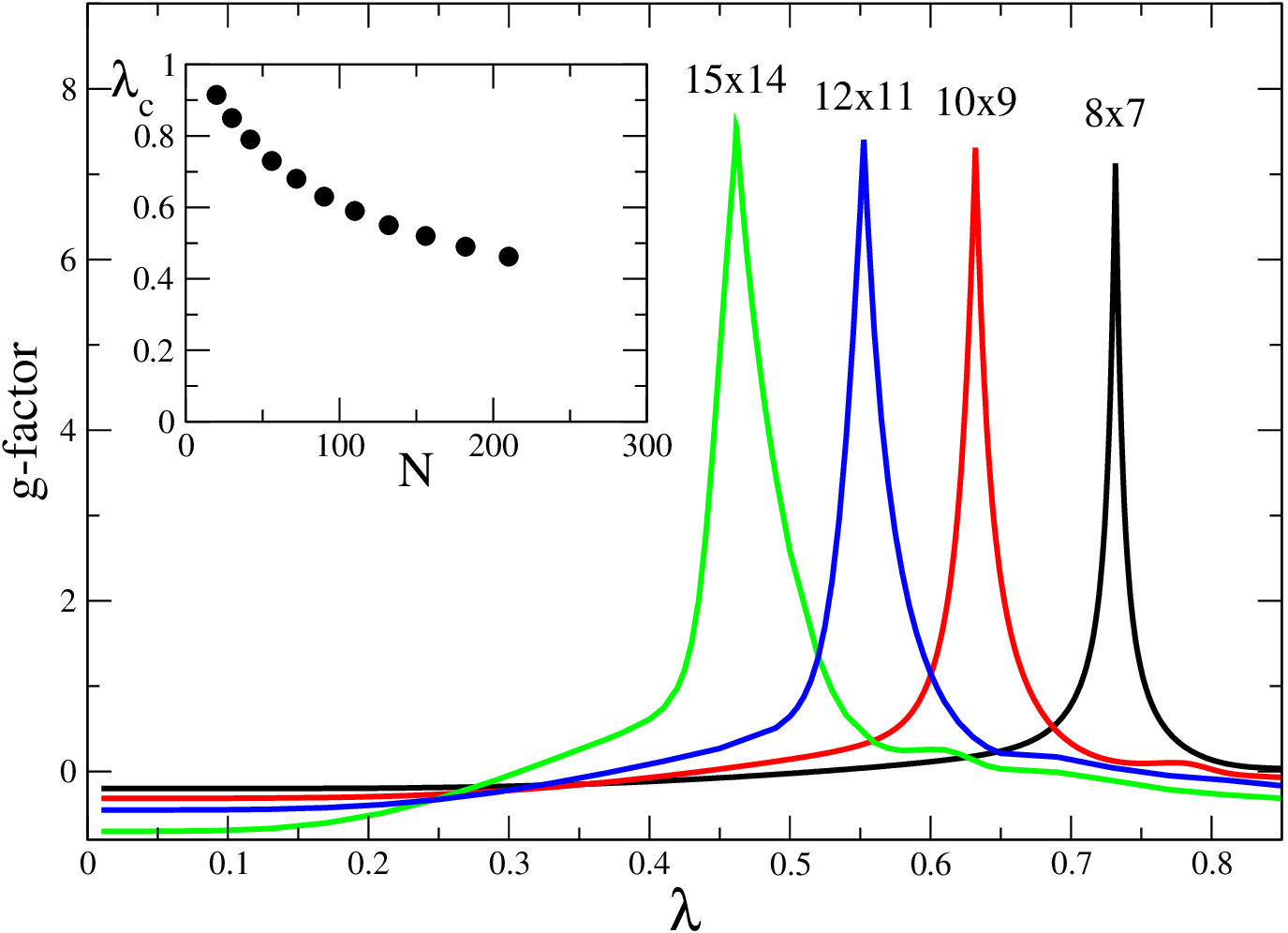}
\caption[The g-factor peak for various system sizes]
{\label{fig:g_so_UCNN} 
(Color online) The g-factor of the doubly-occupied ground-state as a function of the SO coupling strength $\lambda$,
for lattices sizes of $8 \times 7$, $10 \times 9$, $12 \times 11$ and
$15 \times 14$ (size increases from right to left), with $U=3t$. 
Inset: the dependence of $\lambda_c$ on the number of sites.
}
\end{figure}

In Fig.~\ref{fig:g_so_UCNN} we show the dependence of the g-factor on $\lambda$,
for various system sizes, ranging from $8 \times 7$ to $15 \times 14$. 
As can be seen, the enhancement of the g-factor occurs for all of the curves,
with some quantitative changes in the position and the height of the peak.
Although a substantial enlargement of the system is not numerically possible
because of the limitations of the exact diagonalization technique, 
yet the trend is clearly seen.
The value of $\lambda_c$ which is found to decrease with increasing system size
(see the inset of Fig.~\ref{fig:g_so_UCNN})
suggests that for a sufficiently large system size
the crossing occurs for a moderate value of the SO coupling,
which may be experimentally relevant.
In addition, the modest increase of the peak height suggests that
a significant peak may be observed for realistic system sizes.

Finally, we would like to
discuss the implications of the g-factor peak on g-factor measurements.
Once a finite magnetization of the doubly occupied ground state is possible, it can
affect experiments done by, e.g., tunneling spectroscopy. 
Such a measurement presents the result for the difference
between two g-factors, as given by Eq.~(\ref{eqn:g_gdef_tot}). 
If the even-electron state has a non-vanishing g-factor, like in the vicinity of the 
LCs we have presented, the measured quantity $\tilde g$ may not be equal to the 
g-factor of the state with an odd number of electrons, to which it is usually attributed.

\begin{figure}[htbp]
\vskip 1truecm
\centering
\includegraphics[width=3in,height=!]{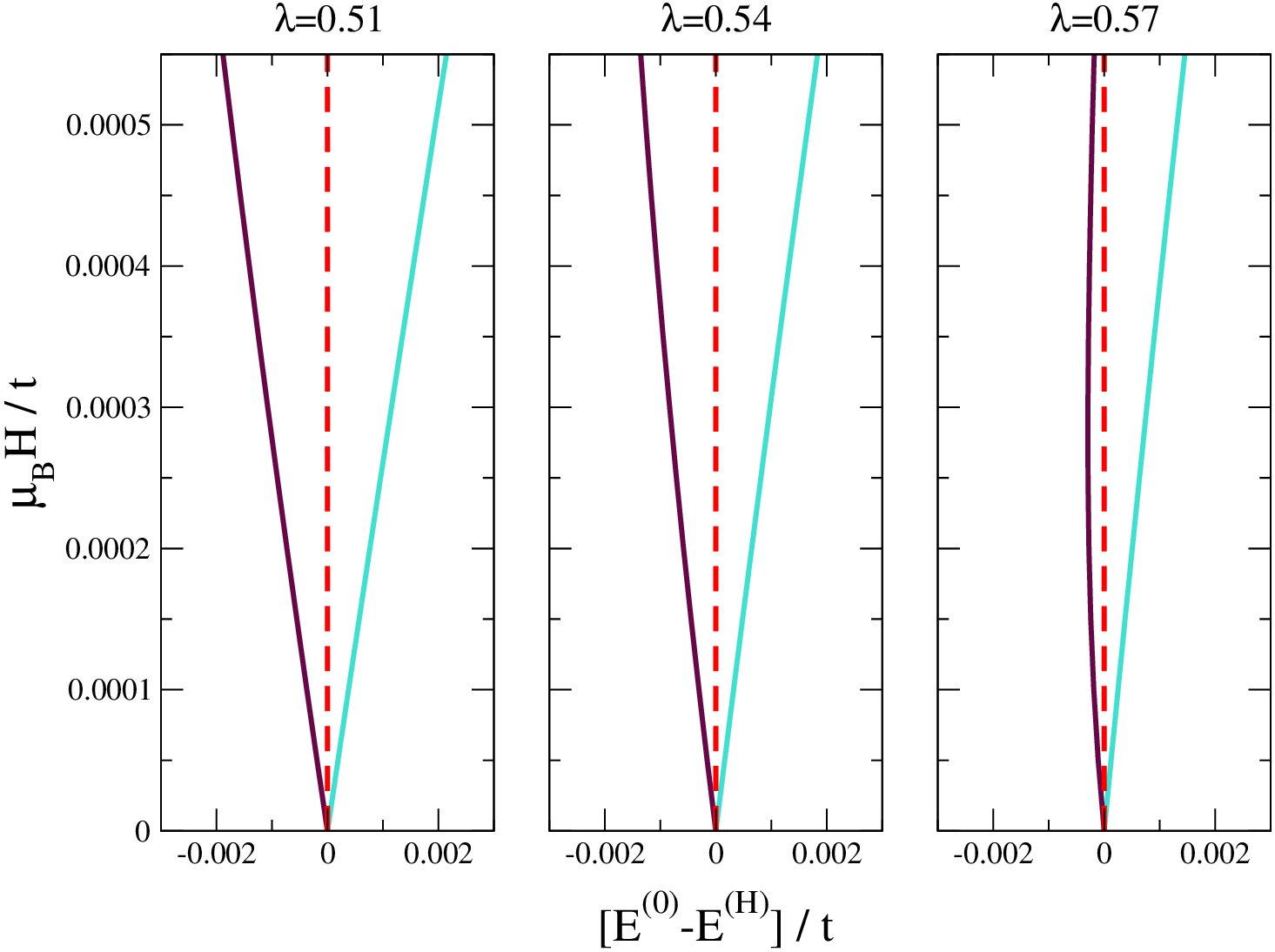}
\caption[Motion of Coulomb peaks with magnetic field]
{\label{fig:E0_conc} 
(Color online) The magnetic field dependence of the first Coulomb peaks ($n_e=1,2$) 
for a lattice of $11 \times 10$ with $U=3t$ (for which
$\lambda_c \approx 0.59$).
}
\end{figure}

In such cases, a signature of the LC may be seen experimentally. In the
regular case (as opposed to the LC scenario), the two levels which belong to the same 
Kramers doublet have the same g-factor up to a sign, and the dependence of the two 
energies on the magnetic field is symmetric. However, in the region of a LC, the two levels receive 
contributions from different even-particle states.
Explicitly, with a magnetic field, the $p$-th Kramers pair is split to levels with
different g-factors, $g(2p-1)-g(2p-2)$ and $g(2p)-g(2p-1)$. 
Thus, if $g(2p)$ or $g(2p-2)$ (or both) are not negligible,
the magnetic field dependence will not be symmetric. 
Furthermore, since the strength of the spin-orbit coupling can be tuned by using
a gate-voltage\cite{nitta97,engels97}, different shapes of the magnetic-field dependence 
may be obtained for a specific sample with different values of the gate-voltage.
An example is presented in Fig.~\ref{fig:E0_conc}.
As one can see, the clearest non-symmetric behavior is obtained for $\lambda \approx \lambda_c$
(right panel), but such a dependence can be seen for a region in its vicinity as well (middle panel). 
Far enough from this region (left panel) the symmetric dependence reappears.

We have also tried to verify that the reported phenomenon occurs for states with even 
electron numbers larger than $2$ as well.
However, the treatment of such cases is more difficult 
since the size of the Hilbert space, $\binom{2AB}{n_e}$, quickly passes
the computational limit when $n_e$ increases. Therefore, the numerical
simulation is limited to much smaller lattices, and although they show features
similar to LC and enhanced g-factors reported for the two electron ground state, 
the question whether
a LC occurs for larger lattices as well needs further investigation.
As a possible method for that calculation we suggest the particle-hole version of the
density matrix renormalization group algorithm\cite{yw_phdmrg}, 
which is suitable for such finite Fermi systems\cite{phdmrg}.
In addition, since as mentioned above the interplay between interactions
and disorder can also result in a non-trivial spin polarization,
the combination of both disorder and SO coupling with interactions can enhance 
this finding. These two points deserve a separate investigation.

To conclude, we have shown that the combination of interactions and spin-orbit 
scattering can lead to a magnetization of states having 
an even number of electrons. This effect was explained using first order
perturbation theory by the tendency of interactions to drive a Stoner 
instability, which is enhanced by the SO coupling.
By studying the behavior when the system size increases, it seems 
that such a result may be experimentally observed even for realistic sizes of QDs.
Therefore it might be relevant for understanding g-factor measurements.
Based on our explanation of the results,
we believe that similar phenomena might be observed in metallic
nano-particles as well. However, a numerical investigation for 3D 
systems is quite difficult.

\acknowledgments

Support from the Israel Academy of Science 
(Grant 569/07) is gratefully acknowledged. 
M.G. is supported by the Adams Fellowship program of 
the Israel Academy of Sciences and Humanities.

\end{document}